# Taking the Plunge: Nanoscale Chemical Imaging of Functionalized Gold Triangles in $H_2O$ *via* TERS

*Ashish Bhattarai,[1] Alan G. Joly,[1] Andrey Krayev,[2] and Patrick Z. El-Khoury[1,*]*

[1]Physical Sciences Division, Pacific Northwest National Laboratory, P.O. Box 999, Richland, WA 99352, USA

[2]Horiba Instruments Inc., 359 Bel Marin Keys Blvd, Suite 20, Novato, CA 94949, USA

[*]patrick.elkhoury@pnnl.gov

**ABSTRACT**

We demonstrate fast (0.25 s/pixel) nanoscale chemical imaging in aqueous solution *via* tip-enhanced Raman scattering (TERS), with sub-15 nm spatial resolution. The substrate consists of 4-mercaptobenzonitrile-functionalized monocrystalline gold triangular platelets immersed in $H_2O$, which we map using a gold-coated AFM probe illuminated using a 633 nm laser source. We find that the recorded TERS images trace the enhanced local optical fields that are sustained towards the edges and corners of the triangles. In effect, we directly map local optical fields of a plasmonic substrate in aqueous solution through molecular TERS. Overall, our described platform and approach may generally be used for chemical and biological imaging, and potentially, to follow chemical transformations at solid-liquid interfaces through TERS.

## Introduction

Single nanometer precision in ambient nanoscale chemical imaging is attainable through tip-enhanced Raman scattering (TERS),[1,2,3] a technique that combines plasmon-enhanced Raman scattering with scanning probe microscopy.[4,5,6,7] Very recently, direct nucleic acid sequencing was achieved using TERS,[3] which not only bolster sub-1 nm spatial resolution but also establishes single base detection sensitivity under ambient laboratory conditions. Even higher spatial resolution has been previously demonstrated in single molecule TERS measurements that take advantage of the stability afforded by ultra-high vacuum and cryogenic temperatures.[8,9,10] Indeed, these works together pave the way for the evolution of TERS from a chemical physicist's tool to a "chemist's microscope".[11] Beyond the realm of atomically-resolved measurements, TERS is continuously shedding new light on fundamental (bio)molecular processes of relevance to fields as diverse as structural biology, catalysis and energy conversion.[2,11,12,13] Nonetheless, other than a few attempts over the past decade that are summarized in a recent review,[13] liquid-phase TERS measurements are scarce. This is somewhat surprising, as many catalytic, electrochemical, as well as biological processes[14] of general interest naturally occur at solid-liquid interfaces.[13]

The first attempt at performing TERS-based characterization of a solid-liquid interface appeared a decade ago.[15] The pioneering proof-of-principle work was only revisited in 2015,[16] whereby the benefits of performing TERS in solution were systematically illustrated. For example, the solution environment was found to shield the scanning probe from exposure to air and impurities. It was also found to act as a heat sink that prevents abrupt local temperature changes in the vicinity of the plasmonic tip upon resonant excitation.[16] More recently, three attempts were made to characterize liquid-solid interfaces *via* TERS. Two scanning tunneling microscopy-based

TERS studies probed resonant[17]/non-resonant[17,18] molecular reporters at interfaces formed between an organic liquid[17] or water[18] on a gold substrate. Another report of relevance to this study described TERS probes that may be used to record ‘tip-only’ atomic force microscopy (AFM)-based TERS in solution, which attests to a high degree of optical field localization and enhancement using the previously described plasmonic tips.[19] The authors further recorded spectrally resolved TERS images of carbon nanotubes in $H_2O$ and inferred a spatial resolution in the 22-62 nm range.[19] Herein, we build on and expand the scope of the aforementioned TERS studies of liquid-solid interfaces. Namely, we describe fast (0.25 s/pixel) and robust spatio-spectral TERS mapping of a 4-mercaptobenzonitrile (MBN)-functionalized monocrystalline gold triangular microplatelets in water and demonstrate sub-15 nm spatial resolution using our scheme.

**Methods**

1:1 gold nanoplates (AP1-10/1000-CTAB-DIH-1-5, NanoPartz™) stock and 1 mM ethanolic 4-mercaptobenzonitrile (MBN, Aurum Pharmatech) solutions were allowed to react for an hour. The mixture is then spin-casted onto a glass bottom solution cell (Cellvis), followed by sonication of the cell in ethanol for ~30 seconds and rigorously washing the substrate using the same solvent. Washing and sonication in ethanol rids the substrate of excess MBN molecules, e.g., unbound reporters on the gold microplates. The alcohol is allowed to dry prior to adding 150 μl of Millipore water for the liquid TERS measurements that are described below.

Our AFM/TERS setup is described elsewhere.[1,20] For the purpose of this work, initial topographic AFM measurements were performed in tapping mode feedback using a silicon tip (Opus, 160AC-NN) coated with 100 nm of Au by arc-discharge physical vapor deposition (target: Ted Pella Inc., 99.99% purity). A 633 nm laser (100-200 μW) is incident onto the apex of the TERS probe using a 100X air objective (Nikon, NA=0.85) using the bottom (inverted) excitation channel. The polarization of the laser is controlled with a half waveplate and is orthogonal to the long axis of the AFM probe (in-plane). The scattered radiation is collected through the same objective and filtered through a series of filters. The resulting light is detected by a CCD camera (Andor, Newton EMCCD) coupled to a spectrometer (Andor, Shamrock 500). A dedicated TERS imaging mode (SpecTop, Patent # US9,568,495B2) was employed for simultaneous AFM-TERS mapping in $H_2O$. Using this mode, TERS signals are collected when the tip is in direct contact with the surface with a typical force in the 10−25 nN range. A semicontact mode is then used to move the sample relative to the tip (pixel to pixel) to preserve the sharpness and optical properties of the tip and to minimize the lateral forces that otherwise perturb the substrate.

FDTD simulations were performed using a commercially available software package (Lumerical Inc.) running on a local computer cluster. The computational models used replicate our experimental geometry by accounting for sample permittivity, laser wavelength, polarization, and angle of incidence. The calculations incorporate a triangular gold nanoplate atop a glass substrate parsed in a three dimensional simulation volume. The calculations yield the spatially resolved relative intensities of the electric field components as a function of time. Standard Fourier transforms result in the corresponding spatial and frequency resolved fields. For computational convenience, a top illumination geometry (in the laboratory frame) is used in our FDTD simulations. This explains some of the differences between the simulated and experimental field profiles, whereby a bottom excitation-collection geometry is used to record the images.

## Results and Discussion

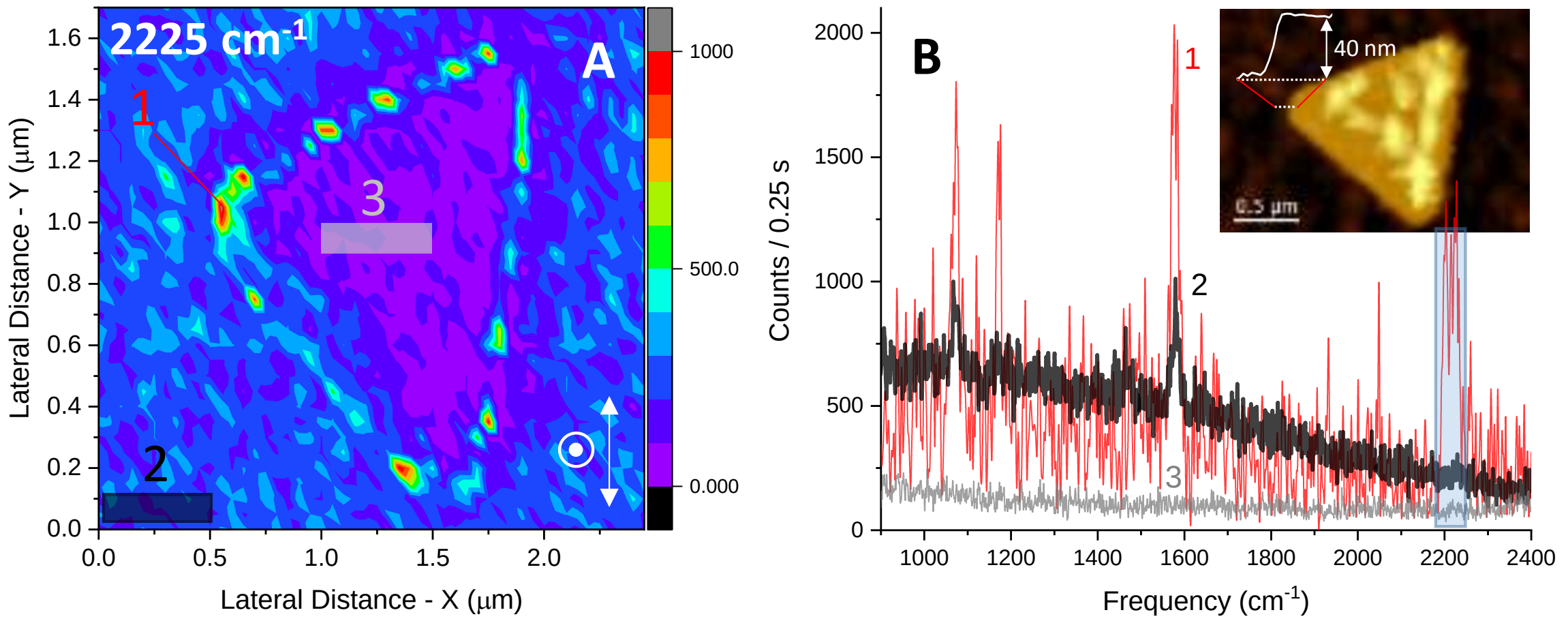

**Figure 1**. (A) A coarse TERS map of an MBN-coated triangular microplate at 2225cm$^{-1}$. The incident light direction and polarization are indicated in the lower-left of the image. Also shown are three regions of interest at which a single pixel spectrum (1) and 2 spatially averaged spectra over the rectangular regions indicated (2 and 3) were extracted. The resulting spectra are shown in (B), the inset of which also shows the simultaneously recorded AFM image of the triangle. Scale bar in (A) corresponds to counts / 0.25 s.

A coarse TERS image/overview of the functionalized gold microplatelets is shown in Figure 1. The gold crystal is supported by a glass coverslip, and a bottom illumination/collection scheme with in-plane polarized 633-nm irradiation (along Y, as indicated in the inset of Figure 1A) is used to record the spectral image. Figure 1A shows a 2.5 μm x 1.7 μm TERS image recorded at ~2225 cm$^{-1}$, near the resonance maximum of the nitrile stretching vibration of MBN.[1] The hyperspectral TERS map from which this image slice is taken was recorded using a (coarse) 50 nm lateral step size, and the signal was integrated for 0.25 s at each pixel. As better illustrated by our ensuing results, the edges of the triangle are selectively visualized in the recorded TERS maps. Selected spectra plotted in Figure 1B also illustrate the concept. These single pixel/spatially averaged spectra were taken from regions of interest that are indicated in Figure 1A. Namely, a single pixel spectrum near the corner of the triangle (region 1) is at least 2X more intense than the tip-only response (spatially averaged over region 2). The tip signal arises from MBN molecules transferred from the substrate to the tip upon landing of the AFM.[1] This is commonly observed when metallic TERS probes are brought into contact with different substrates (e.g., Au, Ag, and even dielectrics) functionalized with aromatic thiols and other (resonant) Raman reporters. The liquid environment does not seem to affect the process. Note that the optical response of the chemically functionalized tip does not change throughout the scan, see Figure S4. Conversely, no measurable response is observed when the tip is on top of the inner part of the gold crystal (region 3). This is because both the incident and scattered fields are attenuated by the 40 nm-thick gold crystal, whereby only ~0.5% of the incident and scattered radiation can be registered on the detector using our bottom excitation/collection geometry. Prior to an exposition of higher spatial resolution TERS mapping, we note that we will exclusively analyze TERS images near the nitrile stretching frequency from hereon. This is in part because the background tails off in the higher frequency region of our spectral region of interest, which results in higher TERS contrast without background subtraction. Moreover, the TERS spectra at the edge feature a much more pronounced nitrile signature, which in turn seems to be suppressed in the 'tip-only' spectra that are recorded when the probe is positioned above the glass support (e.g., region 2 in Figure 2A). The effect is somewhat noticeable in Figure 1B and more evident through full spatio-spectral analysis of the same spectral image in Figure S1.

Figure 2 shows a higher-resolution TERS image (0.75 μm x 1.75 μm, 10 nm lateral step size) that provides the basis for understanding the recorded spatial profiles. Figures 2A and 2B show simultaneously recorded AFM and TERS (at ~2225 $cm^{-1}$) maps. The TERS map image recorded using X-polarized laser illumination selectively exposes the right edge of the triangle, see Figure 2B. This is reminiscent of our recently reported TERS images of MBN-functionalized silver nanowires, whereby the local optical fields of the plasmonic nanostructure were visualized through molecular TERS.[20] Note that unlike our present report, our previous work probed a solid-air interface and utilized a side-illumination geometry.[20] More generally, it is possible to image the nature,[21,22] magnitude,[1,23] and vector components[24] of local optical fields sustained on plasmonic nanostructures through molecular TERS. With this in mind, we performed finite-difference time-domain (FDTD) simulations that illustrate the structure of the local optical fields upon 633 nm laser irradiation of our gold micro-triangular plate, see Figure 2C-F. A comparison between the recorded and simulated local field profiles allows us to preclude the in-plane (X and Y) components, even though the incident polarization is along X (see the inset of Figure 2B). On the basis of this simple side-to-side comparison, it appears that our TERS images broadcast either the total enhanced local optical field or its z-component. The same conclusion is reached when the TERS images of a corner of the triangular gold microplate are compared to its enhanced local fields in Figure S2. In effect, the latter-mentioned measurements can be used to preclude the contribution of the y-component of the local optical field to the structure of the TERS map.

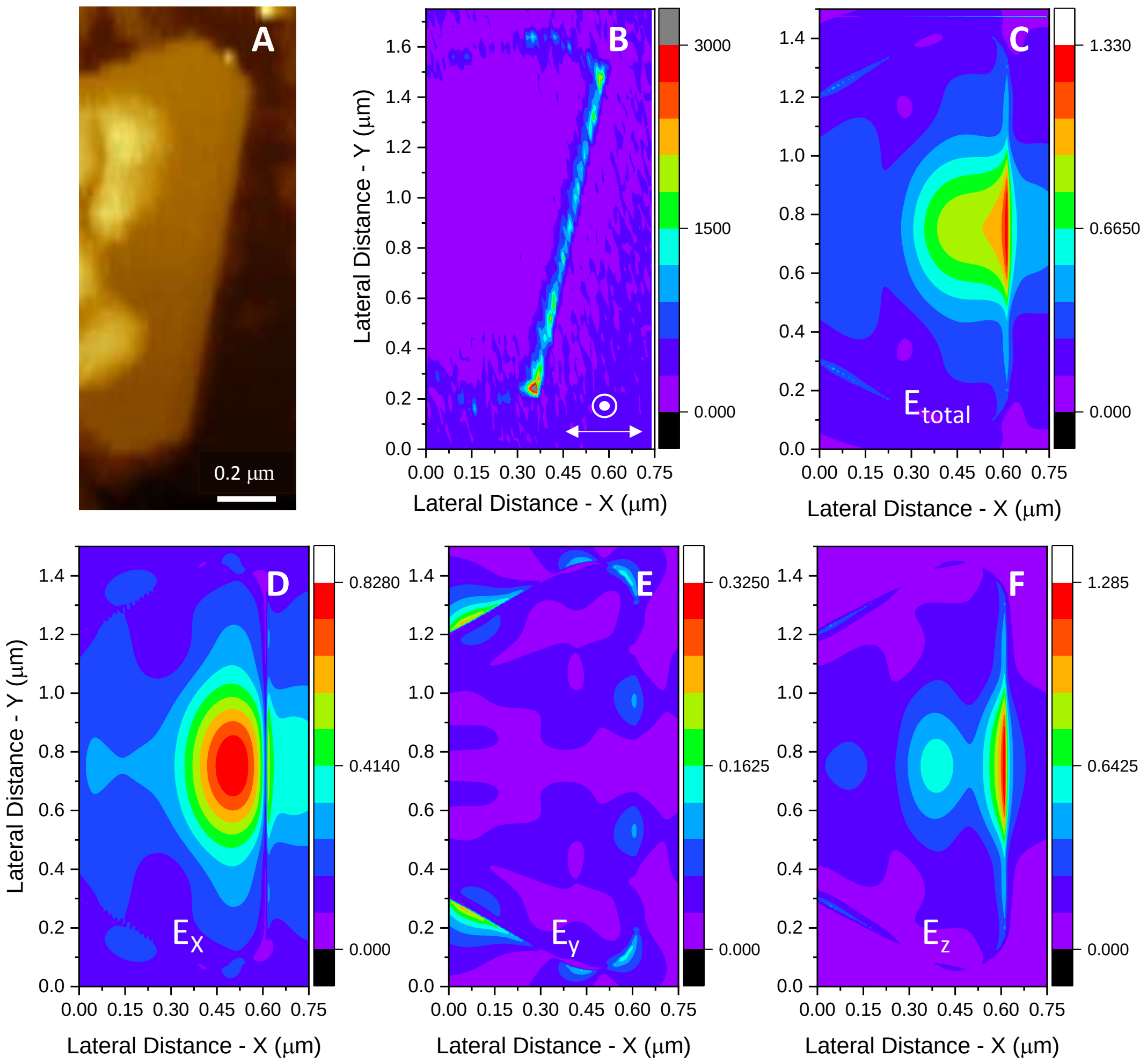

**Figure 2**. Simultaneously recorded AFM (A) and TERS (B) maps (~2225 $cm^{-1}$) are shown. The inset of (B) indicates the direction and polarization of the incident radiation. The TERS image is compared to FDTD simulations of the total (C) and vector components (D-E) of the local optical fields that are sustained following 633 nm irradiation of the triangular gold microplate. Note that the simulated field magnitudes are shown throughout (C-F). The scale bar in (B) indicated counts / 0.25 s.

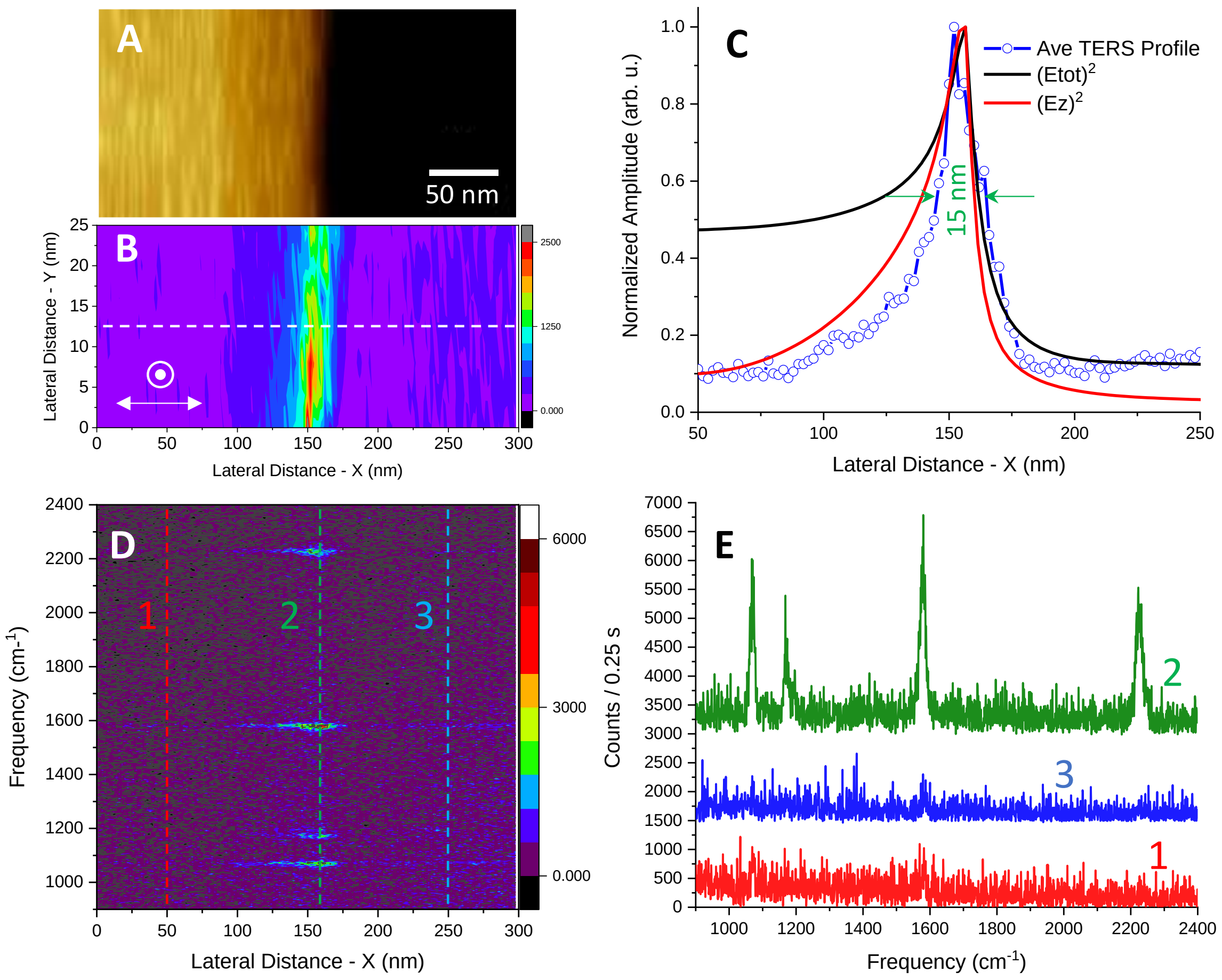

**Figure 3**. Simultaneously recorded AFM (A) and TERS (B) maps are shown. The inset of (B) indicates the direction and polarization of the incident radiation. Spatial averaging (along Y) of the TERS image in (B) results in the average TERS profile shown in (C) along with corresponding numerical profiles taken from Figures 2C and 2F. Spatio-spectral variations along the dashed line (indicated in B) are visualized in (D). In turn, spectral cuts taken at the enumerated positions marked in (D) are plotted in (E). The reader is referred to the main text for more details. Scale bars in (B) and (D) indicate counts / 0.25 s.

In Figure 3, we further zoom into a selected area (300 nm x 25 nm, 2 nm lateral step size) of the same edge that was visualized in Figure 2. The simultaneously recorded AFM and TERS maps are shown in Figures 3A and 3B, respectively. Spatial averaging of the ~2225 $cm^{-1}$ image along the Y-direction allows us to obtain a profile along X (orthogonal to the edge, see Figure 2B) with an improved signal-to-noise ratio, which we then compare to slices of the total/z-component of the simulated local optical field profiles on the same scale. The plots are shown together in Figure 3C. They strongly suggest that the recorded images broadcasts the z-component of the local fields at the edge of the triangular microplate. Notice that a better overall agreement is observed outside of the plate as compared to atop of the gold structure, whereby the experimental field profile decays faster than its simulated analogue. This is attributed to differences between the experimental (bottom excitation/collection) and FDTD (top excitation) geometries. Namely, the incident and scattered fields are attenuated through the gold structure in practice. That the z-component of the local field dominates the overall structure of the TERS maps is consistent with TERS selection rules.[20,24,25,26] That the measured profile closely resembles the square of the z-component of the local field is also in line with prior reports,[24,26] where deviations from the $(E/E_0)^4$ law were documented. In our present work, this observation can be rationalized by recognizing that the in-plane incident polarization (perpendicular to the tip axis) is not enhanced; only the scattered fields that are along the tip axis are broadcasted through the recorded TERS spectral images.

Full spatio-spectral analysis of the TERS image shown in Figure 3B was subsequently performed to closely examine spectral variations across the edge at the different vibrational resonances of MBN.[1] Figure 3D illustrates this analysis. The plot shows that all the vibrational resonances clearly mark the position of the edge with high contrast when small enough lateral steps are used. All the encountered resonances also indicate a field-profile limited sub 15-nm spatial resolution, as shown in Figure 3C for the nitrile stretching vibration. Note that the attainable spatial resolution is only limited by the gently varying spatial profile of the local optical field, as illustrated in Figure 3C. Selected spectra taken at the positions marked by color-coded dashed lines in Figure 3D are offset (by 1500 counts/0.25s) and shown on the same graph in Figure 3E. At least an order of magnitude enhancement of the TERS signal at the edge is observed in Figure 3E, as compared to its analogues that were recorded when the tip was atop the Au structure (position 1 in Figure 3D) or in contact with the glass substrate (position 3 in Figure 3D). The frequency-dependent images otherwise only reflect the relative intensities of the vibrational lines, which is expected in our case of ensemble-averaged TERS spectral imaging.[20,24]

**Conclusions**

In conclusion, this work describes a platform that may be used for fast (0.25 s/pixel), reproducible, and robust TERS in $H_2O$ with sub-15 nm spatial resolution. Reproducibility is ensured by virtue of the symmetry of the triangular target used herein, whereby n-scans across the X-direction (see Figure 2) essentially corresponds to repeating the same measurement n-times. The robustness of the approach is ensured by using gold triangular microplates and a gold-coated tip that were largely unscathed throughout the course of our measurements, as evidenced by reproducible TERS signals across the images. The recorded spectral images trace the z-component of the enhanced local optical field in the vicinity of the edges of the triangular micro-plate, which is consistent with our recent measurements and analyses.[10,20] Admittedly, the use of bottom illumination/collection and in-plane polarization herein is somewhat restrictive and limits the attainable TERS enhancement factors to ~$(E_z/E_0)^2$. This may be improved through the use of light polarized along the Z-direction of light propagation, which is currently being implemented in our lab.

## AUTHOR INFORMATION

### Corresponding Author

*patrick.elkhoury@pnnl.gov

The authors declare no competing financial interest.

**Supporting Information Available:** Spatio-spectral analysis of the coarse TERS map shown in Figure 1, TERS image of a vertex, Helium Ion image of the Au-coated TERS probe, and an expanded analysis of the tip response in Figure 1.

## ACKNOWLEDGMENTS

AB and PZE were supported by the Department of Energy's (DOE) Office of Biological and Environmental Research Bioimaging Technology project #69212. AGJ was supported by the US DOE, Office of Science, Office of Basic Energy Sciences, Division of Chemical Sciences, Geosciences & Biosciences. AK acknowledges support through Horiba Scientific. This work was performed in the environmental and molecular sciences laboratory (EMSL), a DOE Office of Science User Facility sponsored by BER and located at PNNL. PNNL is operated by Battelle

Memorial Institute for the DOE under contract number DE-AC05-76RL1830. The Authors are most grateful to several of their colleagues from Horiba Scientific without whom this would not have been possible, in particular Eddy Robinson and Andrew Whitley.

# TOC Graphic

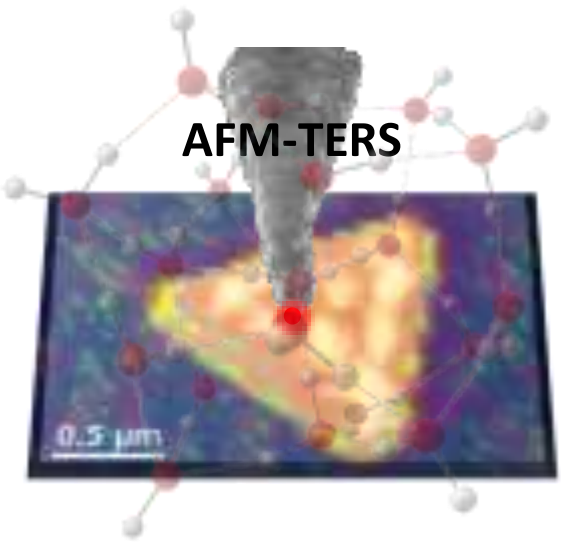